\documentclass[letterpaper, 10 pt, conference]{ieeeconf}
\IEEEoverridecommandlockouts
\let\labelindent\relax
\usepackage{enumitem}
\usepackage{balance}
\usepackage[font={small}]{caption}
\usepackage{subcaption}
\usepackage{array}
\usepackage{textcomp}
\usepackage{mathtools, nccmath}
\usepackage{graphicx}
\usepackage{amsfonts}
\usepackage{amsmath}
\usepackage{amssymb}
\usepackage{algorithm}
\usepackage{algorithmic}
\usepackage{hyperref}
\usepackage{tikz}
\usepackage{arydshln}
\usepackage{multirow}
\usepackage{bm}
\usepackage{epstopdf}
\usepackage{cite}
\usepackage{siunitx}
\usepackage{enumitem}

\usepackage{amsthm}

\newtheorem{theorem}{Theorem}

\newtheorem{remark}{Remark}
\theoremstyle{definition}
\newtheorem{definition}{Definition}
\newtheorem{problem}{Problem}

\title{\LARGE \bf
Model Predictive Control for Safety-Critical Systems \\ using Taylor's Theorem with Lagrange Remainder
}

\author{Shuo Liu$^{1}$, Liang Wu$^{2}$, Wei Xiao$^{3}$, J\'an Drgo\v na$^2$ and Calin A. Belta$^{4}$
\thanks{This work was supported in part by the NSF under grant IIS-2024606
at Boston University and by a Brendan Iribe endowed professorship at the
University of Maryland.}
\thanks{$^{1}$S. Liu is with the Department of Mechanical Engineering, Boston
University, Brookline, MA, USA. 
        {\tt\small liushuo@bu.edu}}%
\thanks{$^{2}$L. Wu and J. Drgona are with the Department of Civil and Systems Engineering, Johns Hopkins University, Baltimore, MD, USA 
        {\tt\small \{wliang14, jdrgona1\}@jh.edu}}
\thanks{$^{3}$W. Xiao is with the School of Electrical and Electronic Engineering, Nanyang Technological University, Singapore and MIT CSAIL, MA, USA.
        {\tt\small weixy@mit.edu}}%
\thanks{$^{4}$C. Belta is with the Department of Electrical and Computer Engineering and the Department of Computer Science, University of Maryland, College Park, MD, USA. 
        {\tt\small calin@umd.edu}}%
}

\begin{document} 
\maketitle
\begin{abstract}
Safety-critical Model Predictive Control (MPC) formulations based on Control Barrier Functions (CBFs) often require multiple tuning parameters and may become conservative or difficult to keep feasible, particularly for high-relative-degree constraints. To address these limitations, in this paper we draw inspiration from the recently introduced Taylor-Lagrange Control (TLC) method, and propose an MPC formulation based on a Taylor expansion with Lagrange remainder (MPC-TLR). To enable digital implementation, we develop a numerical approximation of the Lagrange remainder under zero-order-hold control. The resulting safety constraint includes a tunable margin to account for approximation errors and balance safety, feasibility, and conservativeness. Comparisons with MPC using high-order CBFs (MPC-HOCBF), discrete-time high-order CBFs (MPC-DHOCBF), and direct safety constraints (MPC-DC) on a unicycle obstacle-avoidance problem show that MPC-TLR yields safer trajectories than MPC-DHOCBF and MPC-DC, while requiring fewer tuning parameters and being less conservative than MPC-HOCBF. Its broader control-input coverage under the adopted discretization can improve feasibility and control performance for short prediction horizons.
\end{abstract}

\section{Introduction}
\label{sec:Introduction}
Safety-critical optimal control is a fundamental problem in many areas. For example, in robotics, reaching a target while avoiding obstacles and minimizing control effort can be formulated as a constrained optimal control problem using control barrier functions (CBFs)~\cite{ames2014control,ames2016control}. By discretizing time into small intervals, the control problem can be addressed by solving a quadratic program at each sampling instant, often in real time. Extensions of this framework have been developed for high-relative-degree constraints and adaptive control settings~\cite{nguyen2016exponential,xiao2021high,xiao2021adaptive,liu2023auxiliary,liu2024auxiliary}. However, because each optimization is typically myopic and does not explicitly account for future system behavior, the resulting control actions can be overly aggressive.

Model predictive control (MPC) with discrete-time CBFs (DCBFs)~\cite{zeng2021safety2,agrawal2017discrete} addresses this limitation by incorporating predicted future states over a receding horizon, thereby enabling smoother and more anticipatory control actions. However, the resulting optimization problem is generally nonlinear and nonconvex, and its computational cost may increase substantially with the prediction horizon. Maintaining feasibility is another important challenge. For relative-degree-one DCBFs, relaxation techniques were introduced in~\cite{zeng2021enhancing} to improve feasibility. For higher-relative-degree constraints, an iterative convex MPC framework based on linearized discrete-time high-order CBFs (DHOCBFs)~\cite{xiong2022discrete} was developed in~\cite{liu2025learning}, significantly reducing the computational burden while incorporating relaxation techniques to improve feasibility.

Despite these developments, several fundamental issues remain. 
MPC-CBF formulations, including their discrete-time and high-order extensions, have been widely applied to safety-critical control problems~\cite{sabouni2024reinforcement,yin2023shield,liu2025flexible}. 
However, the validity of the candidate barrier function is not always explicitly established in these formulations. Consequently, the theoretical safety guarantees associated with a valid CBF may not directly carry over. Conversely, if a valid CBF has already been constructed and can be enforced by a feasible controller \cite{zhao2023convex, jiang2026compatible}, MPC is not required to establish safety itself, although it may still improve performance through prediction and optimization. This raises the question of whether an unverified CBF constraint offers a systematic advantage over other forms of safety constraints, and more generally, which safety formulation is preferable within MPC. In addition, high-order CBF formulations introduce more design parameters as the relative degree increases, making it increasingly difficult to balance safety, feasibility, and control conservativeness.

The recently proposed Taylor--Lagrange Control (TLC) method~\cite{xiao2025taylor} ensures system safety by leveraging Taylor's theorem with the Lagrange remainder~\cite{taylor1717methodus,de1813theorie}. Unlike CBF-based approaches, TLC provides a \emph{necessary and sufficient} safety condition while introducing substantially fewer tuning parameters. Subsequent extensions have addressed inter-sampling safety, namely, possible safety violations between consecutive sampling instants under zero-order-hold control, as well as adaptive control settings~\cite{xiao2026robust,liu2026event}. Motivated by these developments, we develop an MPC safety formulation based on a Taylor expansion with a numerically approximated Lagrange remainder. Specifically, the main contributions of this paper are as follows:

\begin{itemize}
    \item We develop a numerically implementable formulation based on a Taylor expansion with Lagrange remainder (TLR) under zero-order-hold control by approximating the integral remainder using intermediate states and numerical quadrature. The number of quadrature points can be adjusted to improve the approximation accuracy.

    \item We integrate the proposed TLR-based formulation into a receding-horizon MPC framework for safety-critical control. A tunable safety margin is introduced to compensate for numerical approximation errors and to adjust the trade-off among safety, feasibility, and control conservativeness.

    \item We show that MPC-TLR provides a good balance among safety, feasibility, and conservativeness, achieving safer trajectories than MPC-DHOCBF and MPC-DC while requiring fewer tuning parameters and being less conservative than MPC-HOCBF. A unicycle obstacle-avoidance example is used to validate these advantages and compare the four formulations under different horizons and parameter settings.
\end{itemize}

\section{Preliminaries}
\label{sec:Preliminaries}

In this section, we review continuous-time High-Order Control Barrier Functions (HOCBFs) and their discrete-time counterpart, discrete-time HOCBFs (DHOCBFs), which will serve as baseline safety formulations for comparisons.

\subsection{High-Order CBF (HOCBF)}
\label{subsec:HOCBF}
Consider an affine control system of the form
\begin{equation}
\label{eq:affine-control-system}
\dot{\boldsymbol{x}}=f(\boldsymbol{x})+g(\boldsymbol{x})\boldsymbol{u},
\end{equation}
 where $\boldsymbol{x}\in \mathbb{R}^{n}, f:\mathbb{R}^{n}\to\mathbb{R}^{n}$ and $g:\mathbb{R}^{n}\to\mathbb{R}^{n\times q}$ are locally Lipschitz, and $\boldsymbol{u}\in \mathcal U\subset \mathbb{R}^{q}$, where $\mathcal U$ denotes the control limitation set, which is assumed to be in the form: 
\begin{equation}
\label{eq:control-constraint}
\mathcal U \coloneqq \{\boldsymbol{u}\in \mathbb{R}^{q}:\boldsymbol{u}_{min}\le \boldsymbol{u} \le \boldsymbol{u}_{max} \}, 
\end{equation}
with $\boldsymbol{u}_{min},\boldsymbol{u}_{max}\in \mathbb{R}^{q}$ (vector inequalities are interpreted componentwise). We assume that no component of $\boldsymbol{u}_{min}$ and $\boldsymbol{u}_{max}$ can be infinite.

In this paper, a \textbf{safety requirement} is defined as $h(\boldsymbol{x})\geq 0$, and \textbf{safety} is the forward invariance of the set $\mathcal C\coloneqq\{\boldsymbol{x}\in\mathbb{R}^{n}:h(\boldsymbol{x})\geq 0\}$. For \(h(\boldsymbol{x})\geq 0\) with continuous-time relative degree \(m\) with respect to system \eqref{eq:affine-control-system} as defined in \cite[Def.~6]{xiao2021high}, where $h:\mathbb{R}^{n}\to\mathbb{R}$ and $\psi_{0}(\boldsymbol{x})\coloneqq h(\boldsymbol{x})$, we define a sequence of functions $\psi_{i}:\mathbb{R}^{n}\to\mathbb{R}$, $i\in\{1,\ldots,m\}$, as

\begin{equation}
\label{eq:sequence-f1}
\psi_{i}(\boldsymbol{x})\coloneqq\dot{\psi}_{i-1}(\boldsymbol{x})+\alpha_{i}(\psi_{i-1}(\boldsymbol{x})),\ i\in \{1,...,m\}, 
\end{equation}
where $\alpha_{i}(\cdot ),\ i\in \{1,...,m\}$ denotes a $(m-i)^{th}$ order differentiable class $\mathcal{K}$ function. We further set up a sequence of sets $\mathcal C_{i}$ based on \eqref{eq:sequence-f1} as
\begin{equation}
\label{eq:sequence-set1}
\mathcal C_{i}\coloneqq \{\boldsymbol{x}\in\mathbb{R}^{n}:\psi_{i}(\boldsymbol{x})\ge 0\}, \ i\in \{0,...,m-1\}. 
\end{equation}
\begin{definition}[HOCBF~\cite{xiao2021high}]
\label{def:HOCBF}
Let $\psi_{i}(\boldsymbol{x}),\ i\in \{1,...,m\}$ be defined by \eqref{eq:sequence-f1} and $\mathcal C_{i},\ i\in \{0,...,m-1\}$ be defined by \eqref{eq:sequence-set1}. A function $h:\mathbb{R}^{n}\to\mathbb{R}$ is a High-Order Control Barrier Function (HOCBF) with relative degree $m$ for system \eqref{eq:affine-control-system} if there exist $(m-i)^{th}$ order differentiable class $\mathcal{K}$ functions $\alpha_{i},\ i\in \{1,...,m\}$ such that
\begin{equation}
\label{eq:highest-HOCBF}
\begin{split}
\sup_{\boldsymbol{u}\in \mathcal U}[L_{f}^{m}h(\boldsymbol{x})+L_{g}L_{f}^{m-1}h(\boldsymbol{x})\boldsymbol{u}+O(h(\boldsymbol{x}))
+\\
\alpha_{m}(\psi_{m-1}(\boldsymbol{x}))]\ge 0,
\end{split}
\end{equation}
$\forall \boldsymbol{x}\in \mathcal C_{0}\cap...\cap \mathcal C_{m-1},$ where $L_f^m h$ and $L_gL_f^{m-1}h$ denote the corresponding Lie derivatives, and
$O(\cdot)=\sum_{i=1}^{m-1}L_{f}^{i}(\alpha_{m-1}\circ\psi_{m-i-1})(\boldsymbol{x})$
collects the remaining Lie-derivative terms. By the relative-degree assumption, $L_gL_f^{m-1}h(\boldsymbol{x})\neq0$ in the domain of interest.
\end{definition}

\begin{theorem}[Safety Guarantee~\cite{xiao2021high}]
\label{thm:safety-guarantee}
Given a HOCBF $h(\boldsymbol{x})$ from Def. \ref{def:HOCBF} with corresponding sets $\mathcal{C}_{0}, \dots,\mathcal {C}_{m-1}$ defined by \eqref{eq:sequence-set1}, if $\boldsymbol{x}(0) \in \mathcal {C}_{0}\cap \dots \cap \mathcal {C}_{m-1},$ then any Lipschitz controller $\boldsymbol{u}$ that satisfies the inequality in \eqref{eq:highest-HOCBF}, $\forall t\ge 0$ renders $\mathcal {C}_{0}\cap \dots \cap \mathcal {C}_{m-1}$ forward invariant for system \eqref{eq:affine-control-system}, $i.e., \boldsymbol{x} \in \mathcal {C}_{0}\cap \dots \cap \mathcal {C}_{m-1}, \forall t\ge 0.$
\end{theorem}

\subsection{Discrete-time HOCBF (DHOCBF)}
\label{subsec:DHOCBF}
Consider an affine control system of the form \eqref{eq:affine-control-system}. By applying a numerical discretization
scheme to \eqref{eq:affine-control-system}, such as the forward Euler method \cite{hairer1993solving}, we obtain the discrete-time system
\begin{equation}
\label{eq:discrete-dynamics}
\boldsymbol{x}_{t+1}
=
F(\boldsymbol{x}_{t},\boldsymbol{u}_{t}),
\end{equation}
where $t\in\mathbb{N}$ and
$F:\mathbb{R}^{n}\times\mathbb{R}^{q}\to\mathbb{R}^{n}$ denotes the
resulting discrete-time state-transition map. We assume that $h(\boldsymbol{x})$ has
relative degree $m_d$ with respect to system \eqref{eq:discrete-dynamics} based on [\text{Def}. 1, \cite{liu2025learning}]. Informally, the discrete-time relative degree $m_d$ is the minimum number of steps required for $\boldsymbol{u}_t$ to affect $h(\boldsymbol{x}_{t+m_d})$. The continuous-time relative degree counts time differentiations, whereas $m_d$ counts discrete-time steps and may depend on the discretization scheme. Starting with $\psi_{0}(\boldsymbol{x}_{t})\coloneqq h(\boldsymbol{x}_{t})$, we define a sequence of discrete-time functions $\psi_{i}:  \mathbb{R}^{n}\to\mathbb{R}$, $i=1,\dots,m_d$ as:
\begin{equation}
\label{eq:high-order-discrete-CBFs}
\psi_{i}(\boldsymbol{x}_{t})\coloneqq \bigtriangleup \psi_{i-1}(\boldsymbol{x}_{t})+\alpha_{i}(\psi_{i-1}(\boldsymbol{x}_{t})), 
\end{equation}
where $\bigtriangleup \psi_{i-1}(\boldsymbol{x}_{t})\coloneqq \psi_{i-1}(\boldsymbol{x}_{t+1})-\psi_{i-1}(\boldsymbol{x}_{t})$, and $\alpha_{i}(\cdot)$ denotes the $i^\text{th}$ class $\mathcal{K}$ function which satisfies $\alpha_{i}(\psi_{i-1}(\boldsymbol{x}_{t}))\le \psi_{i-1}(\boldsymbol{x}_{t})$ for $i=1,\ldots, m_d$. A sequence of sets $\mathcal C_i$ is similarly defined as 
\eqref{eq:sequence-set1} for $i\in \{0,...,m_d-1\}$. 
\begin{definition}[DHOCBF~\cite{xiong2022discrete}]
\label{def:high-order-discrete-CBFs}
Let $\psi_{i}(\boldsymbol{x}), \ i\in \{1,\dots,m_d\}$ be defined by \eqref{eq:high-order-discrete-CBFs} and $\mathcal {C}_{i},\ i\in \{0,\dots,m_d-1\}$ be defined by \eqref{eq:sequence-set1}. A function $h:\mathbb{R}^{n}\to\mathbb{R}$ is a Discrete-time High-Order Control Barrier Function (DHOCBF) with relative degree $m_d$ for system \eqref{eq:discrete-dynamics} if there exist $\psi_{m_d}(\boldsymbol{x})$ and $\mathcal {C}_{i}$ such that
\begin{equation}
\label{eq:highest-order-CBF}
\psi_{m_d}(\boldsymbol{x}_{t})\ge 0, \ \forall \boldsymbol{x}_{t}\in \mathcal{C}_{0}\cap \dots \cap \mathcal {C}_{m_d-1}, t\in\mathbb{N}.
\end{equation}
\end{definition}

\begin{theorem}[Safety Guarantee \cite{xiong2022discrete}]
\label{thm:forward-invariance}
Given a DHOCBF $h(\boldsymbol{x})$ from Def. \ref{def:high-order-discrete-CBFs} with corresponding sets $\mathcal{C}_{0}, \dots,\mathcal {C}_{m_d-1}$ defined by \eqref{eq:sequence-f1}, if $\boldsymbol{x}_{0} \in \mathcal {C}_{0}\cap \dots \cap \mathcal {C}_{m_d-1},$ then any Lipschitz controller $\boldsymbol{u}_{t}$ that satisfies the constraint in \eqref{eq:highest-order-CBF}, $\forall t\ge 0$ renders $\mathcal {C}_{0}\cap \dots \cap \mathcal {C}_{m_d-1}$ forward invariant for system \eqref{eq:discrete-dynamics}, $i.e., \boldsymbol{x}_{t} \in \mathcal {C}_{0}\cap \dots \cap \mathcal {C}_{m_d-1}, \forall t\ge 0.$
\end{theorem}

When their respective relative degrees are one, HOCBF and DHOCBF reduce to the (zeroing) control barrier function (CBF) \cite{ames2016control} and discrete-time control barrier function (DCBF) \cite{agrawal2017discrete}, respectively. Both provide sufficient rather than necessary conditions for safety. For the HOCBF formulation in \eqref{eq:sequence-f1}, $\alpha_i(0)=0$ implies that $\psi_i\geq0$ reduces to $\dot{\psi}_{i-1}\geq0$ on the boundary $\psi_{i-1}=0$, preventing the trajectory from leaving the corresponding set. Enforcing this inequality in the interior is stronger than the boundary condition required for forward invariance. Similarly, the DHOCBF formulation in \eqref{eq:high-order-discrete-CBFs} implies $\psi_{i-1}(\boldsymbol{x}_{t+1})\geq\psi_{i-1}(\boldsymbol{x}_t)-\alpha_i(\psi_{i-1}(\boldsymbol{x}_t))\geq0$, given $\psi_{i-1}(\boldsymbol{x}_t)\geq0$ and $\alpha_i(\psi_{i-1}(\boldsymbol{x}_t))\leq \psi_{i-1}(\boldsymbol{x}_t)$. This is stronger than directly requiring $\psi_{i-1}(\boldsymbol{x}_{t+1})\geq0$. Therefore, a trajectory may remain safe while violating the prescribed HOCBF or DHOCBF inequalities.

\section{Problem Formulation and Approach}
\label{sec:Problem Formulation and Approach}
Our objective is to design a closed-loop control policy for system \eqref{eq:affine-control-system} over the time interval $[0,T]$ that drives the system toward a desired state while ensuring safety and satisfying state and control input constraints.

\textbf{Safety Requirement:}
The continuous-time system \eqref{eq:affine-control-system} is required to satisfy
\begin{equation}
\label{eq:Safety_constraint}
h(\boldsymbol{x}(t)) \geq 0,
\quad \forall t\in[0,T],
\end{equation}
where $h:\mathbb{R}^{n}\to\mathbb{R}$ is a continuously differentiable
function.

\textbf{Control and State Constraints:}
The state and control input must satisfy
\begin{equation}
\boldsymbol{x}(t)\in\mathcal{X},
\quad
\boldsymbol{u}(t)\in\mathcal{U},
\quad
\forall t\in[0,T].
\end{equation}

\textbf{Objective:}
We consider the finite-time optimal control objective
\begin{equation}
\label{eq:continuous-cost}
J
=
\int_{0}^{T}
\left(
\|\boldsymbol{x}(t)-\boldsymbol{x}_d\|_{Q}^{2}
+
\|\boldsymbol{u}(t)\|_{R}^{2}
\right)\,dt
+
\|\boldsymbol{x}(T)-\boldsymbol{x}_d\|_{P}^{2},
\end{equation}
where $\boldsymbol{x}_d$ denotes the desired state, $Q\succeq0$, $R\succ0$, and $P\succeq0$ are weighting matrices ($\|\boldsymbol{z}\|_{W}^{2}\coloneqq\boldsymbol{z}^{\top}W\boldsymbol{z}$ denotes a weighted quadratic form).
\begin{problem}
\label{prob:Path-prob}
Find a feasible control policy for system \eqref{eq:affine-control-system} such that the safety requirement, control and state limitations are satisfied, and cost \eqref{eq:continuous-cost} is minimized.
\end{problem}

Conventional CBF controllers compute the current input through a one-step optimization. For a fixed state $\boldsymbol{x}$, $f(\boldsymbol{x})$ and $g(\boldsymbol{x})$ in \eqref{eq:affine-control-system} are fixed, so the CBF and HOCBF conditions are affine in $\boldsymbol{u}$. With a quadratic objective, they can therefore be imposed as linear constraints in a quadratic program (QP)~\cite{nguyen2016exponential,xiao2021high}. Because only the current input is optimized, these controllers do not account for future state evolution and may produce myopic actions.

MPC addresses this limitation by optimizing a sequence of control inputs over a finite prediction horizon while explicitly accounting for the predicted state evolution. It has therefore been integrated with DCBFs~\cite{zeng2021safety2} and DHOCBFs~\cite{liu2023iterative}. In particular, it was shown in~\cite{zeng2021safety2} that MPC-DCBF can increase the minimum distance to obstacles compared with directly imposing $h(\boldsymbol{x})\geq 0$ as a state constraint. To formulate these predictive controllers, the continuous-time dynamics are discretized with a sampling period $\Delta t$, and the continuous-time cost in \eqref{eq:continuous-cost} is approximated by a finite sum over the prediction horizon.

Nevertheless, the HOCBF and DHOCBF conditions provide sufficient rather than necessary conditions for safety as discussed in Sec. \ref{subsec:DHOCBF}, and their class $\mathcal{K}$ functions introduce additional tuning parameters that may affect feasibility and conservativeness. In contrast, TLC provides necessary and sufficient conditions for satisfying the safety requirement while requiring fewer tuning parameters~\cite{xiao2025taylor,liu2026event}. However, existing TLC formulations do not explicitly account for predicted future states. Motivated by this limitation, we develop a predictive safety formulation based on a Taylor expansion with Lagrange remainder (TLR) and incorporate it into an MPC framework. We refer to the resulting approach as MPC-TLR and compare it with existing safety-critical MPC approaches in the following sections.

\textbf{Approach:} In the proposed approach, the continuous-time dynamics \eqref{eq:affine-control-system} are first discretized with a sampling period $\Delta t$ to construct the discrete-time model \eqref{eq:discrete-dynamics} used in MPC, and the integral cost in \eqref{eq:continuous-cost} is approximated by a finite sum over the prediction horizon. The safety requirement is incorporated into the resulting discrete-time optimization problem through TLR-based constraints imposed on the predicted states. A tunable safety margin is introduced into these constraints to provide an additional safety margin and adjust the conservativeness of the controller. The TLR-based constraints are then integrated with the state and control input bounds, yielding a finite-horizon nonlinear and generally nonconvex optimization problem that is solved at each sampling instant.

\section{Taylor–Lagrange-Based Model Predictive Control}
\label{sec:Taylor–Lagrange-Based Model Predictive Control}
This section incorporates the general $m^\text{th}$-order TLR representation into a receding-horizon MPC framework. The predicted inputs are applied using zero-order hold (ZOH), while intermediate states are introduced to approximate the continuous-time trajectory and the integral remainder term. The resulting formulation combines the necessary and sufficient safety characterization induced by the Taylor--Lagrange representation with finite-horizon prediction, and is compared with state-of-the-art safety-critical MPC formulations based on direct safety constraints, HOCBFs, and DHOCBFs.
\subsection{General $m^\text{th}$-Order Taylor--Lagrange Formulation}
\label{subsec:Taylor--Lagrange Formulation}

Let the safety function $h(\boldsymbol{x})$ have relative degree $m$ with respect to system \eqref{eq:affine-control-system}. For any $t\geq0$ and $\Delta t>0$, applying the fundamental theorem of calculus to $h(\boldsymbol{x}(t))$ gives
\begin{equation}
h\bigl(\boldsymbol{x}(t+\Delta t)\bigr)
=
h\bigl(\boldsymbol{x}(t)\bigr)
+
\int_{t}^{t+\Delta t}
h^{(1)}\bigl(\boldsymbol{x}(t_1)\bigr)\,dt_1.
\end{equation}
Recursively applying the same argument to the derivatives $h^{(i)}$, $i=1,\ldots,m-1$, yields
\begin{equation}
\begin{aligned}
&h\bigl(\boldsymbol{x}(t+\Delta t)\bigr)
={}
\sum_{i=0}^{m-1}
\frac{\Delta t^i}{i!}
h^{(i)}\bigl(\boldsymbol{x}(t)\bigr)+ \\
&
\int_t^{t+\Delta t}
\int_t^{t_1}
\cdots
\int_t^{t_{m-1}}
h^{(m)}
\bigl(\boldsymbol{x}(t_m),\boldsymbol{u}(t_m)\bigr)
\,dt_m\cdots dt_2dt_1.
\end{aligned}
\end{equation}
Here $
h^{(m)}
=
L_f^m h(\boldsymbol{x}(t_m))
+
L_gL_f^{m-1}h(\boldsymbol{x}(t_m))
\boldsymbol{u}(t_m)$ in the integral remainder. Reversing the order of integration in the remainder term yields
\begin{equation}
\label{eq:general_tlc_integral}
\begin{aligned}
&h\bigl(\boldsymbol{x}(t+\Delta t)\bigr)
={}
\sum_{i=0}^{m-1}
\frac{\Delta t^i}{i!}
h^{(i)}\bigl(\boldsymbol{x}(t)\bigr) \\
&+
\int_{t}^{t+\Delta t}
\frac{(t+\Delta t-t_m)^{m-1}}{(m-1)!}
h^{(m)}
\bigl(\boldsymbol{x}(t_m),\boldsymbol{u}(t_m)\bigr)\,dt_m.
\end{aligned}
\end{equation}
The TLR expression \eqref{eq:general_tlc_integral} does not introduce recursive class $\mathcal K$ functions. Since its right-hand side is exactly equal to $h(\boldsymbol{x}(t+\Delta t))$, the safety condition at time $t+\Delta t$ is satisfied if and only if the right-hand side of \eqref{eq:general_tlc_integral} is nonnegative. This expression, however, does not by itself guarantee inter-sample safety over
$(t,t+\Delta t)$. Although \eqref{eq:general_tlc_integral} is exact, evaluating its remainder requires the state $\boldsymbol{x}(t_m)$ at every $t_m\in[t,t+\Delta t]$. Even under ZOH control, this state trajectory is generally unavailable in closed form. Therefore, we approximate the remainder numerically, as detailed in Sec.~\ref{subsec:ZOH Implementation}.
\subsection{ZOH Implementation and Numerical Approximation of TLR}
\label{subsec:ZOH Implementation}
To implement the TLR expression \eqref{eq:general_tlc_integral} in a digital controller, we assume that
the control input is updated at each sampling time $t$ and held
constant over the sampling interval under a ZOH, i.e., $\boldsymbol{u}(t+s)=\boldsymbol{u}(t),
~s\in[0,\Delta t).$
Under the ZOH policy, the integral remainder in
\eqref{eq:general_tlc_integral} becomes
\begin{equation}
\int_{0}^{\Delta t}
\frac{(\Delta t-s)^{m-1}}{(m-1)!}
h^{(m)}
\bigl(\boldsymbol{x}(t+s),\boldsymbol{u}(t)\bigr)\,ds .
\label{eq:zoh-integral-remainder}
\end{equation}
Although the control input is constant over the sampling interval, the
state $\boldsymbol{x}(t+s)$ evolves continuously and is generally not
available in closed form. We therefore introduce $M+1$ intermediate
time points $s_j=\frac{j}{M}\Delta t,~ j=0,\ldots,M$ and denote the corresponding states by
\begin{equation}
\boldsymbol{x}_{t,j}
\approx
\boldsymbol{x}(t+s_j),
\end{equation}
where $\boldsymbol{x}_{t,0}=\boldsymbol{x}(t)$. The intermediate states are obtained by numerically integrating the
continuous-time dynamics by \eqref{eq:discrete-dynamics} over each subinterval while applying the
constant input $\boldsymbol{u}(t)$, i.e., $\boldsymbol{x}_{t,j+1}
=
F(\boldsymbol{x}_{t,j},\boldsymbol{u}(t))$. Using these intermediate states, the integral remainder is approximated
through numerical quadrature as
\begin{equation}
\begin{aligned}
&\int_{0}^{\Delta t}
\frac{(\Delta t-s)^{m-1}}{(m-1)!}
h^{(m)}
\bigl(\boldsymbol{x}(t+s),\boldsymbol{u}(t)\bigr)\,ds \\
&\qquad \approx
\sum_{j=0}^{M}
w_j
\frac{(\Delta t-s_j)^{m-1}}{(m-1)!}
h^{(m)}
\bigl(\boldsymbol{x}_{t,j},\boldsymbol{u}(t)\bigr),
\end{aligned}
\label{eq:numerical-integral-remainder}
\end{equation}
where $w_j$ are the quadrature weights.

\begin{remark}
The accuracy of the numerical approximation \eqref{eq:numerical-integral-remainder} generally improves as the
number of subintervals $M$ increases, provided that the integrand is
sufficiently smooth and a convergent numerical quadrature rule is used.
However, increasing $M$ also increases the computational cost.

To compensate for the numerical approximation error, a safety margin
$\delta\geq 0$ can be introduced by enforcing
\begin{equation}
\label{eq:numerical TLC}
\begin{aligned}
\tilde{h}\bigl(\boldsymbol{x}(t+\Delta t)\bigr)=\sum_{i=0}^{m-1}\frac{{\Delta t}^i}{i!}
h^{(i)}\bigl(\boldsymbol{x}(t)\bigr)+\\\sum_{j=0}^{M}
w_j
\frac{(\Delta t-s_j)^{m-1}}{(m-1)!}
h^{(m)}
\bigl(\boldsymbol{x}_{t,j},\boldsymbol{u}(t)\bigr)
\geq \delta,
\end{aligned}
\end{equation}
where $\tilde{h}\bigl(\boldsymbol{x}(t+\Delta t)\bigr)$ denotes the
future safety-function value obtained using the numerical approximation
of the integral remainder. A larger $\delta$ yields a more conservative
constraint, whereas $\delta=0$ corresponds to the nominal numerical
approximation. In particular, if the approximation error satisfies
\begin{equation}
\left|
h\bigl(\boldsymbol{x}(t+\Delta t)\bigr)
-
\tilde{h}\bigl(\boldsymbol{x}(t+\Delta t)\bigr)
\right|
\leq \varepsilon_M,
\end{equation}
then selecting $\delta\geq\varepsilon_M$ guarantees
$h\bigl(\boldsymbol{x}(t+\Delta t)\bigr)\geq0$.
\end{remark}

\subsection{MPC-TLR formulation}
We now incorporate the numerically implementable TLR condition \eqref{eq:numerical TLC} into a finite-horizon MPC framework. At each time step $t$, the continuous-time integral cost in \eqref{eq:continuous-cost} is approximated by a finite sum over the prediction horizon $N$ with a sampling period $\Delta t$, thereby accounting for the predicted evolution of the system over the next $N$ states:
\noindent\rule{\columnwidth}{0.4pt}
  \textbf{MPC-TLR:}  
\begin{subequations}
\label{eq:mpc-tlc}
\begin{align}
\min_{\mathbf{U}_{t}}
\quad &
\left\|
\boldsymbol{x}_{t+N}-\boldsymbol{x}_{d}
\right\|_{P}^{2}
+
\sum_{k=0}^{N-1}
\left(
\left\|
\boldsymbol{x}_{t+k}-\boldsymbol{x}_{d}
\right\|_{Q}^{2}
+
\left\|
\boldsymbol{u}_{t+k}
\right\|_{R}^{2}
\right)
\label{eq:mpc-tlc-cost}
\\
\mathrm{s.t.}\quad
&
\boldsymbol{x}_{t+0}
=
\boldsymbol{x}(t),
\label{eq:mpc-tlc-initial}
\\
&
\boldsymbol{x}_{t+k+1}
=
F
\left(
\boldsymbol{x}_{t+k},
\boldsymbol{u}_{t+k}
\right),
\quad k=0,\ldots,N-1,
\label{eq:mpc-tlc-dynamics}
\\
&
\sum_{j=0}^{M}
w_j
\frac{(\Delta t-s_j)^{m-1}}{(m-1)!}
h^{(m)}
\bigl(\boldsymbol{x}_{t+k,j},\boldsymbol{u}_{t+k}\bigr)\nonumber
\\&+\sum_{i=0}^{m-1}\frac{{\Delta t}^i}{i!}
h^{(i)}\bigl(\boldsymbol{x}_{t+k}\bigr)\geq \delta,
\quad k=0,\ldots,N-1,
\label{eq:mpc-tlc-constraint}
\\
&
\boldsymbol{x}_{t+k}\in\mathcal{X},
\quad k=0,\ldots,N,
\label{eq:mpc-tlc-state-constraint}
\\
&
\boldsymbol{u}_{t+k}\in\mathcal{U},
\quad k=0,\ldots,N-1,
\label{eq:mpc-tlc-input-constraint}
\end{align}
\end{subequations}
\noindent\rule{\columnwidth}{0.4pt}
where $\boldsymbol{x}_{t+k+1}$ denotes the state predicted $k+1$ steps ahead at time $t$, obtained by applying the control input $\boldsymbol{u}_{t+k}$ to the predicted state $\boldsymbol{x}_{t+k}$ and $\mathbf{U}_{t}$ denotes the control-input sequence over the prediction horizon.
In \eqref{eq:mpc-tlc-cost}, $Q$, $R$, and $P$ denote the state, control input, and terminal cost matrices, respectively. The initial condition and discrete-time prediction dynamics are represented by \eqref{eq:mpc-tlc-initial} and \eqref{eq:mpc-tlc-dynamics}, respectively. The TLR-based safety constraints along the prediction horizon are imposed by \eqref{eq:mpc-tlc-constraint}, while the state and control input constraints are captured by \eqref{eq:mpc-tlc-state-constraint} and \eqref{eq:mpc-tlc-input-constraint}, respectively. The optimal solution to \eqref{eq:mpc-tlc} at time $t$ is $\mathbf{U}_{t}^{*}
=
\left[
\boldsymbol{u}_{t}^{*},
\ldots,
\boldsymbol{u}_{t+N-1}^{*}
\right]$.
The first element of $\mathbf{U}_{t}^{*}$ is applied to the continuous-time system \eqref{eq:affine-control-system} under ZOH as
\begin{equation}
\label{eq: continuous inplementation}
\dot{\boldsymbol{x}}(\tau)
=f(\boldsymbol{x}(\tau))+g(\boldsymbol{x}(\tau))\boldsymbol{u}_{t}^{*},
\quad
\tau\in[t,t+\Delta t),
\end{equation}
to obtain the updated state $\boldsymbol{x}_{t+1}=\boldsymbol{x}(t+\Delta t)$. The constrained finite-horizon optimal control problem \eqref{eq:mpc-tlc} is then solved at time $t+1$, and at all subsequent time steps based on the updated state, yielding an MPC-TLR-based safety-critical receding-horizon control strategy.

The MPC-TLR formulation in \eqref{eq:mpc-tlc} generally results in a nonlinear and nonconvex optimization problem. This nonconvexity arises from the nonlinear prediction dynamics $F(\cdot)$ and the TLR constraint \eqref{eq:mpc-tlc-constraint}, which depends nonlinearly on the predicted states, control inputs, and the $m^\text{th}$-order derivative of the safety function. The computational complexity increases with the prediction horizon $N$, the number of quadrature points $M+1$, and the TLR order $m$. In particular, the TLR constraint requires evaluating $h^{(m)}(\boldsymbol{x}_{t+k,j},\boldsymbol{u}_{t+k})$ at each quadrature point and at every prediction step, resulting in approximately $N(M+1)$ additional nonlinear function evaluations and safety constraints. Moreover, the number of decision variables grows approximately linearly with $N$, while the solution time of a general nonlinear programming solver may grow considerably faster due to the nonconvexity of the problem. Therefore, in practice, \eqref{eq:mpc-tlc} is typically solved using a local nonlinear programming method, such as sequential quadratic programming or an interior-point method, and the obtained solution is generally only locally optimal. 

\subsection{Comparative Analysis with Existing Safety-Critical MPC Formulations}
\label{subsec: Comparative Analysis}

The MPC-TLR formulation shares the same receding-horizon optimization
framework with several existing safety-critical MPC methods. In particular,
the objective function \eqref{eq:mpc-tlc-cost}, initial condition \eqref{eq:mpc-tlc-initial}, prediction dynamics \eqref{eq:mpc-tlc-dynamics}, and state and
input constraints \eqref{eq:mpc-tlc-state-constraint}, \eqref{eq:mpc-tlc-input-constraint} remain unchanged. The main difference lies in the safety
constraint in \eqref{eq:mpc-tlc-constraint}. Therefore, the different methods
can be obtained by replacing \eqref{eq:mpc-tlc-constraint} with their
corresponding safety conditions.

For MPC with direct state constraints, referred to as MPC-DC, safety is
directly imposed on predicted future states as
\begin{equation}
h\left(\boldsymbol{x}_{t+k+1}\right)\geq 0,
\quad k=0,\ldots,N-1.
\label{eq:mpc-dc-constraint}
\end{equation}

For MPC-HOCBF, we rewrite the continuous-time HOCBF condition in \eqref{eq:highest-HOCBF} based on the continuous-time affine control system in \eqref{eq:affine-control-system} and impose the resulting safety constraint at each prediction step as
\begin{equation}
\begin{aligned} 
&L_{f}^{m}h(\boldsymbol{x}_{t+k})
+
L_{g}L_{f}^{m-1}h(\boldsymbol{x}_{t+k})
\boldsymbol{u}_{t+k} +O\left( h\left(\boldsymbol{x}_{t+k}\right) \right) +\\ &\alpha_m\left( \psi_{m-1}\left(\boldsymbol{x}_{t+k}\right) \right) \geq 0, \quad k=0,\ldots,N-1.
\end{aligned}
\label{eq:mpc-hocbf-constraint}
\end{equation}

For MPC-DHOCBF, the discrete-time high-order safety condition derived from \eqref{eq:high-order-discrete-CBFs} is imposed as
\begin{equation}
\psi_{m_d}\left(
\boldsymbol{x}_{t+k}
\right)
\geq 0,
\quad k=0,\ldots,N-m_d.
\label{eq:mpc-dhocbf-constraint}
\end{equation}
The reduced index range in \eqref{eq:mpc-dhocbf-constraint} results from the forward-difference construction in \eqref{eq:high-order-discrete-CBFs}, which requires an $m_d$-step state sequence to formulate the discrete-time high-order condition.

From the above formulations, we can compare the number of control inputs that explicitly enter the safety constraints over the prediction horizon. MPC-TLR and MPC-HOCBF provide the largest input coverage, with up to $N$ control inputs involved in their respective safety constraints. For MPC-DC, the control input affects the safety function $h$ only after $m_d$ prediction steps, where the discrete-time relative degree $m_d$ depends on the chosen discretization scheme. Consequently, when $m_d>1$, only $N-m_d+1$ control inputs explicitly enter the direct state-safety constraints. Similarly, MPC-DHOCBF constructs its safety condition through $m_d$ successive forward differences, so its constraint can only be imposed over the shortened horizon and also involves $N-m_d+1$ control inputs. The intermediate points used in MPC-TLR are introduced solely to numerically approximate the Lagrange remainder under the same ZOH control input and therefore do not introduce additional control decision variables. Safety is still enforced only at the MPC sampling instants, as in the baseline methods; the baseline constraints require no such intermediate points because they contain no integral remainder.

\begin{remark}
\label{rem: 2}
When the discrete-time relative degree $m_d$ of the safety function $h$ is close to the prediction horizon $N$, MPC-TLR and MPC-HOCBF involve substantially more control inputs in their safety constraints than MPC-DC and MPC-DHOCBF. This increased control-input coverage provides the optimizer with greater flexibility to shape the predicted trajectory
for safety and may therefore improve feasibility and reduce
conservativeness. In contrast, when $N$ is significantly larger than $m_d$, the relative difference in control-input coverage becomes small, and its effect on the performance of the four formulations may become negligible.
\end{remark}

Compared with MPC-DC and MPC-DHOCBF, the safety constraints in MPC-TLR and MPC-HOCBF generally have more complex analytical forms. In particular, MPC-TLR requires the evaluation of higher-order derivatives and a numerical approximation of the integral remainder \eqref{eq:mpc-tlc-constraint}, whereas MPC-HOCBF involves successive derivatives of the barrier function \eqref{eq:mpc-hocbf-constraint}. These additional operations
increase the cost of evaluating the constraints and their Jacobians and may therefore increase the computational burden of the nonlinear MPC problem. However, the overall solution time also depends on the number of imposed constraints, their sparsity, the prediction horizon, and the numerical solver,
so the more complex constraint form does not necessarily imply a proportionally larger computation time.

Although the HOCBF, DHOCBF, and DC formulations impose exact safety constraints in their respective formulations, the TLR constraint relies on a numerical approximation of the integral remainder. Nevertheless, the safety of interest is that of the original continuous-time system \eqref{eq:affine-control-system}, whereas all four MPC formulations employ a
discretized prediction model \eqref{eq:mpc-tlc-dynamics} and are therefore subject to discretization error. Moreover, MPC-TLR and MPC-HOCBF derive their safety conditions directly from the continuous-time dynamics \eqref{eq:affine-control-system}, while MPC-DC and MPC-DHOCBF construct them from the discretized dynamics \eqref{eq:mpc-tlc-dynamics}. Consequently, none of the four formulations provides an error-free safety guarantee for the actual continuous-time implementation \eqref{eq: continuous inplementation} unless the discretization error is explicitly accounted for. This discrepancy can be reduced by using a more accurate discretization scheme or a smaller sampling interval $\Delta t$.

\section{Implementation and Results}
\label{sec:Implementation and Results}
In this section, we present numerical results for a unicycle model to evaluate the proposed MPC-TLR approach. We compare its performance with three baseline methods: MPC-HOCBF, MPC-DHOCBF, and MPC-DC.
\subsection{Numerical Setup}
\subsubsection{System Dynamics} Consider a continuous-time unicycle model in the form
\begin{equation}
\label{eq:unicycle-model}
\begin{bmatrix}
\dot{x}(t) \\
\dot{y}(t) \\
\dot{\theta}(t)\\
\dot{v}(t)
\end{bmatrix}  
=\begin{bmatrix}
 v(t)\cos{(\theta(t))}  \\
 v(t)\sin{(\theta(t))} \\
 0 \\
 0
\end{bmatrix} 
+ \begin{bmatrix}
  0 & 0\\
  0 & 0\\
  1 & 0\\
  0 & 1 
\end{bmatrix}\begin{bmatrix}
   u_{1}(t)   \\
  u_{2}(t) 
\end{bmatrix},
\end{equation}
where $\boldsymbol{x}(t)=[x(t),y(t),\theta(t),v(t)]^{\top}$ captures the 2-D location, heading angle, and linear speed; $\boldsymbol{u}(t)=[u_1(t),u_{2}(t)]^{\top}$ represents angular velocity ($u_{1}$) and linear acceleration ($u_{2}$), respectively. The discrete-time model used in this paper is obtained by discretizing
\eqref{eq:unicycle-model} using the forward Euler method with a sampling time of $\Delta t=0.1s$, yielding the dynamics in
\eqref{eq:mpc-tlc-dynamics}. The resulting discrete-time system is subject to the following state and control input constraints:
\begin{equation}
\begin{split}
\label{eq:state-input-constraint}
\mathcal{X}&=\{\boldsymbol{x}_t\in \mathbb{R}^{4}: -10\cdot \mathcal{I}_{4\times1} \le \boldsymbol{x}_t\le 10\cdot \mathcal{I}_{4\times1}\},\\
\mathcal{U}&=\{\boldsymbol{u}_t\in \mathbb{R}^{2}: [-7, -5]^{\top} \le \boldsymbol{u}_{t}\le [7, 5]^{\top}\}.
\end{split}
\end{equation}

\subsubsection{System Configuration}
The initial state is $[-3,0,0,0]^{T}$ and the target state is $\boldsymbol{x}_{e}=[3,0.01,0,0]^{T},$ which are marked as blue and red diamonds in Fig.~\ref{fig:1}. The circular obstacle is centered at $(0,0)$ with $r=1$, which is displayed in grey. We set the $y$-component of the reference state $\mathbf{x}_{d}$ to $0.01$ to avoid singularity in the optimization problem.

\subsubsection{Safety Function and Parameters}
We choose a quadratic distance function as a candidate safety function for circular obstacle avoidance $h(\boldsymbol{x}_{t})= (x_{t}-x_{0})^{2}+(y_{t}-y_{0})^{2}-r^{2}$, where $(x_{0},y_{0})$ and $r$ denote the obstacle center location and radius, respectively. Since the safety function has relative degree $m=2$ with respect to
system~\eqref{eq:unicycle-model}, the HOCBF formulation in \eqref{eq:sequence-f1} contains two class $\mathcal{K}$ functions. Both are chosen to be linear, with parameters $k_1$ and $k_2$. The parameters of MPC-TLR in \eqref{eq:mpc-tlc-constraint} are $\delta$ and $M$, with $M=3$. Similarly, the right-hand-side
parameter in the MPC-DC constraint \eqref{eq:mpc-dc-constraint} can also be denoted by $\delta$. The safety function has discrete-time relative degree $m_d=2$ with respect to the forward-Euler discretized system; therefore, the DHOCBF formulation in \eqref{eq:high-order-discrete-CBFs} contains two
class $\mathcal{K}$ functions, which are chosen to be linear with parameters $\gamma_1$ and $\gamma_2$.

\subsubsection{Cost Function and Parameters} The MPC objective in \eqref{eq:mpc-tlc-cost} consists of stage and terminal costs, with the weighting matrices selected as $Q=P=10\cdot\mathcal{I}_{4}$ and $R=\mathcal{I}_{2}$. To ensure a fair comparison among the four methods, all parameters and environmental settings are kept identical across the experiments.

\subsubsection{Solver Configurations and CPU Specs.}
All approaches were implemented in MATLAB using YALMIP~\cite{lofberg2004yalmip} and solved with IPOPT~\cite{biegler2009large}. The control inputs obtained from IPOPT were applied to the continuous-time system under ZOH, and \texttt{ode45} was used to integrate the system dynamics over each $0.1s$ sampling interval. All computations were performed on a Windows desktop equipped with an Intel Core i7-11700 CPU running at 2.50 GHz.

\subsection{Performance} 
\begin{figure*}[!t]
    \vspace*{-2mm}
    \centering
    \begin{subfigure}[t]{0.245\linewidth}
        \centering
        \includegraphics[width=1\linewidth]{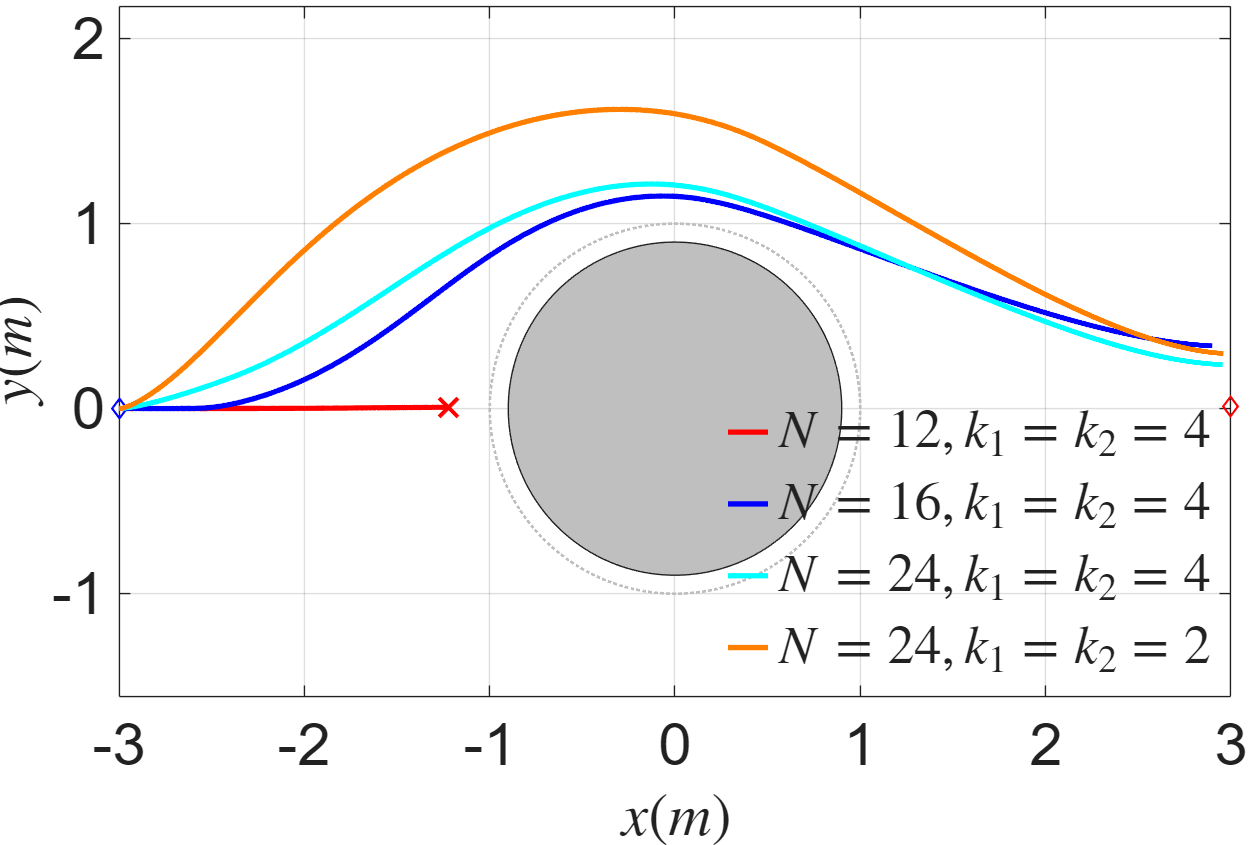}
        \caption{MPC-HOCBF.}
        \label{fig:1a}
    \end{subfigure}
    \begin{subfigure}[t]{0.245\linewidth}
        \centering
        \includegraphics[width=1\linewidth]{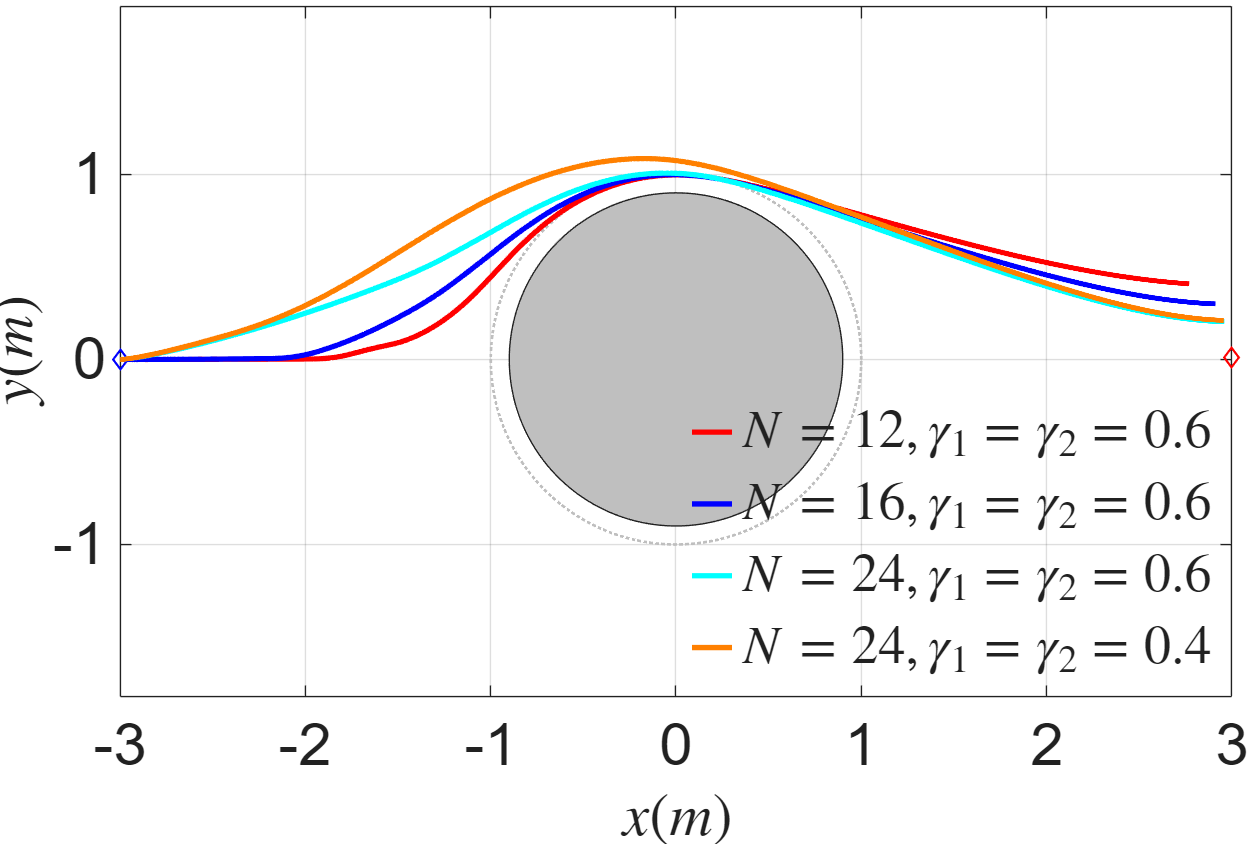}
        \caption{MPC-DHOCBF.}
        \label{fig:1b}
    \end{subfigure}  
    \begin{subfigure}[t]{0.245\linewidth}
        \centering
        \includegraphics[width=1\linewidth]{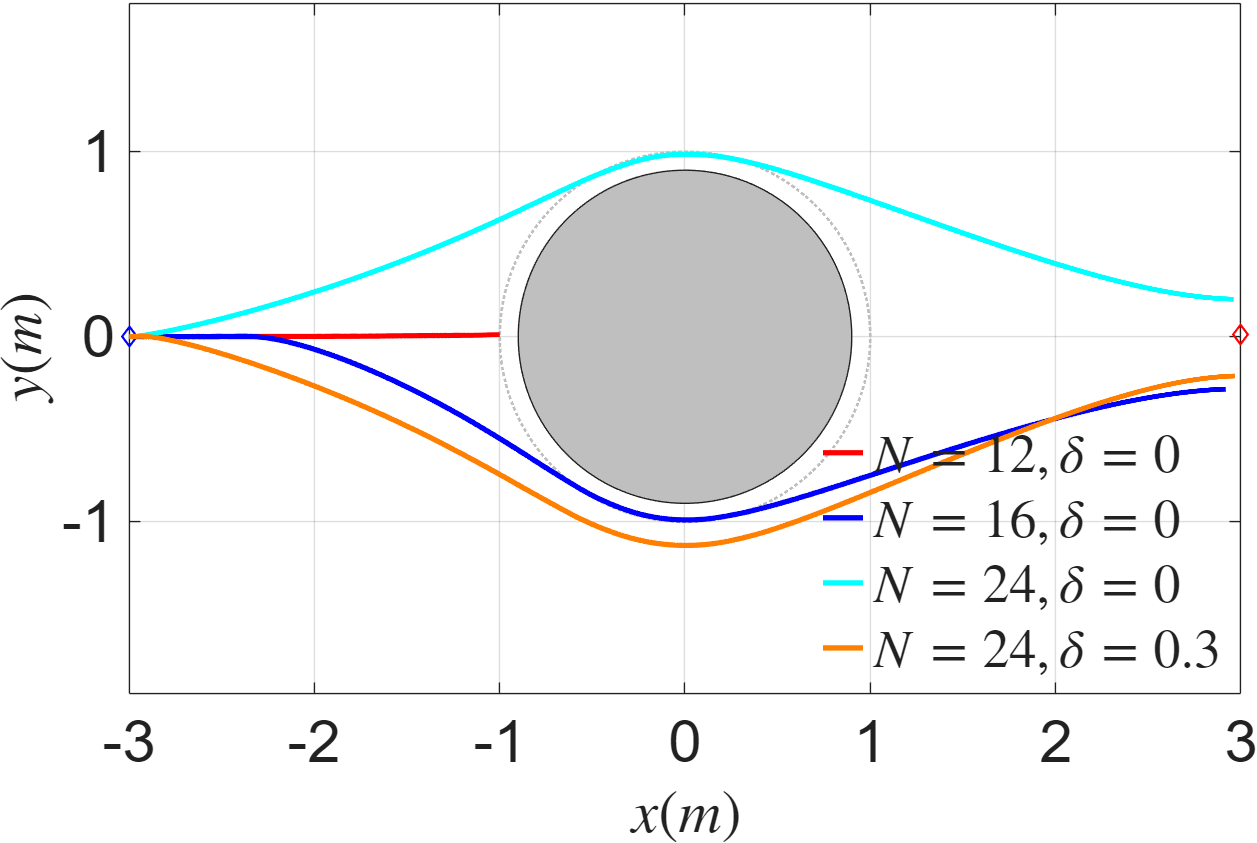}
        \caption{MPC-DC.}
        \label{fig:1c}
    \end{subfigure}
    \begin{subfigure}[t]{0.245\linewidth}
        \centering
        \includegraphics[width=1\linewidth]{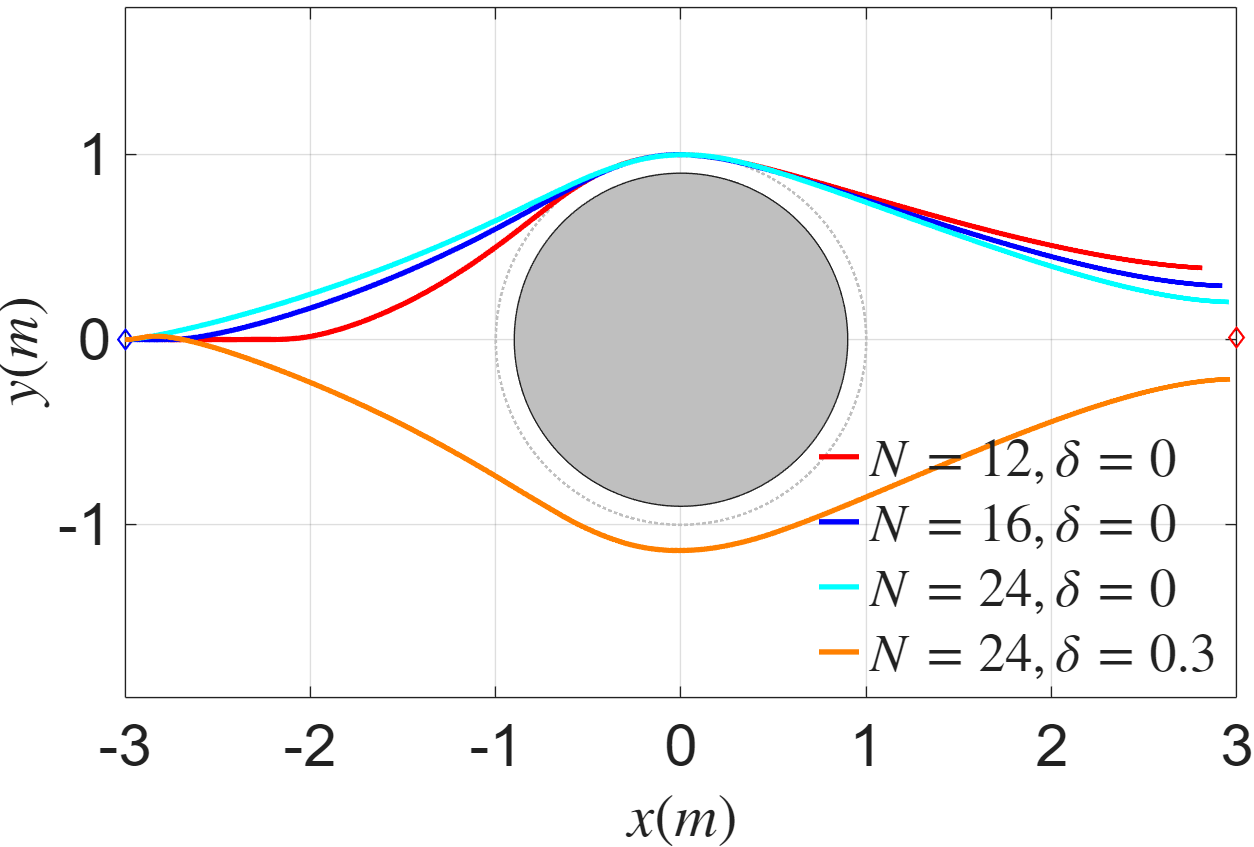}
        \caption{MPC-TLR.}
        \label{fig:1d}
    \end{subfigure}
    \caption{Closed-loop trajectories generated by (a) MPC-HOCBF, (b) MPC-DHOCBF, (c) MPC-DC, and (d) MPC-TLR under different prediction horizons and safety-constraint parameters. The dashed gray circle indicates the actual obstacle boundary, while the filled gray disk is shown slightly smaller for visual clarity.
    } 
    \label{fig:1}
\end{figure*}

\begin{figure*}[!t]
    \vspace*{-0.4cm}
    \centering
    \begin{subfigure}[t]{0.245\linewidth}
        \centering
        \includegraphics[width=1\linewidth]{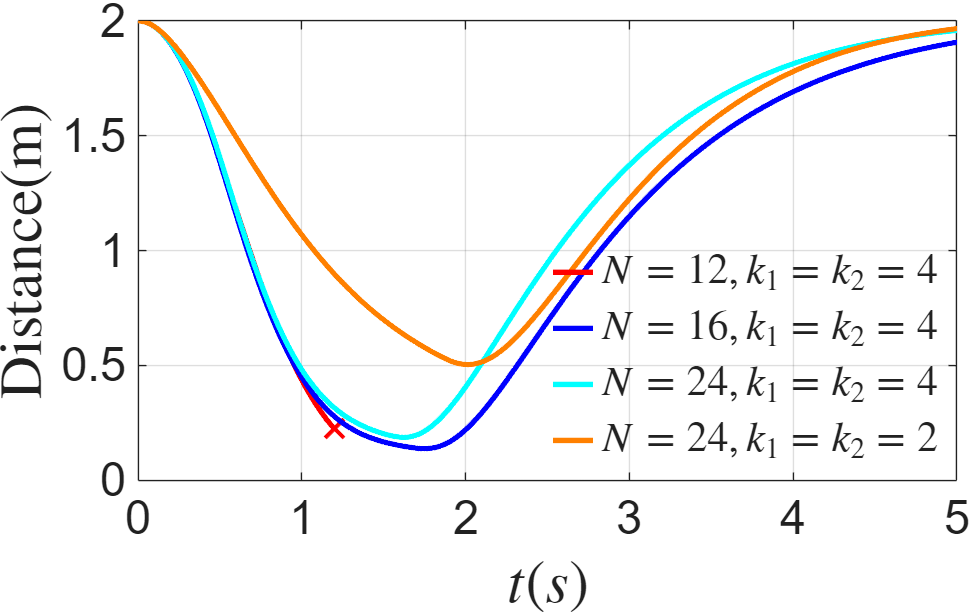}
        \caption{MPC-HOCBF.}
        \label{fig:2a}
    \end{subfigure}
    \begin{subfigure}[t]{0.245\linewidth}
        \centering
        \includegraphics[width=1\linewidth]{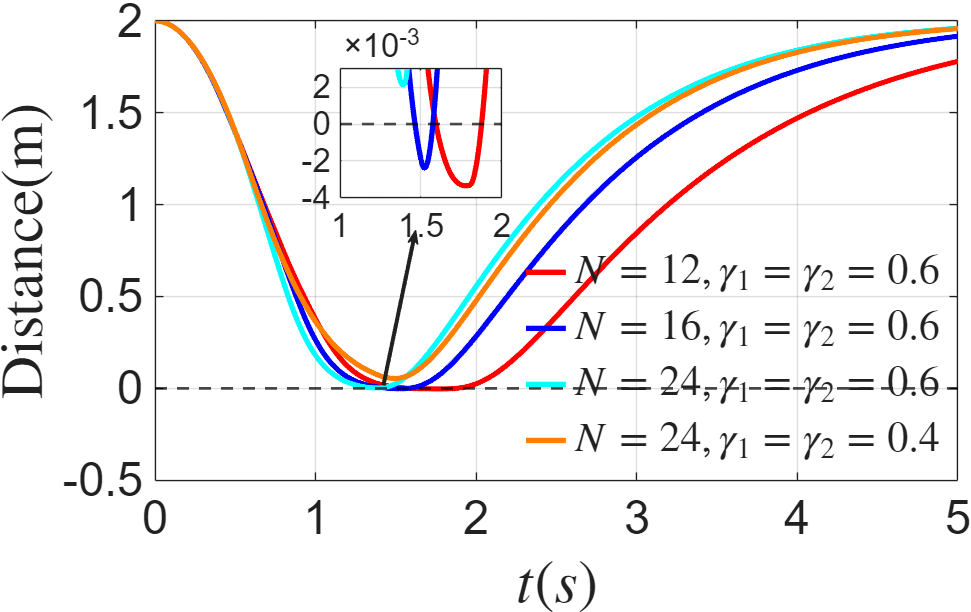}
        \caption{MPC-DHOCBF.}
        \label{fig:2b}
    \end{subfigure}  
    \begin{subfigure}[t]{0.245\linewidth}
        \centering
        \includegraphics[width=1\linewidth]{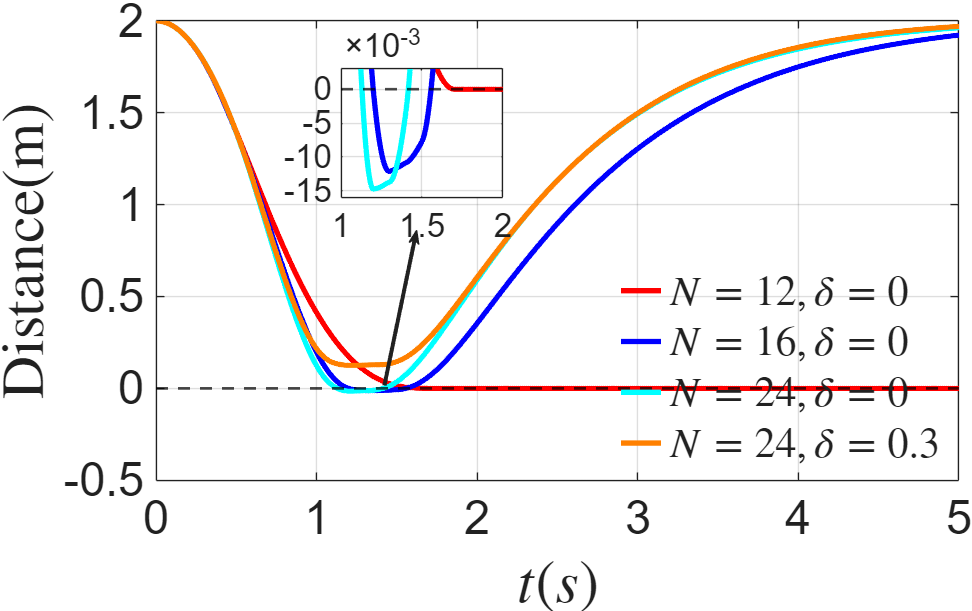}
        \caption{MPC-DC.}
        \label{fig:2c}
    \end{subfigure}
    \begin{subfigure}[t]{0.245\linewidth}
        \centering
        \includegraphics[width=1\linewidth]{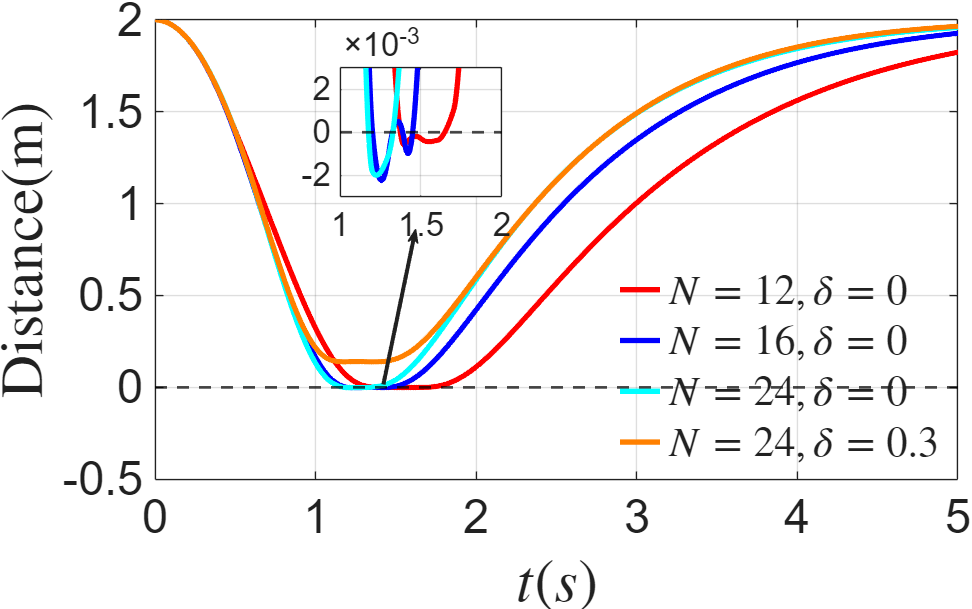}
        \caption{MPC-TLR.}
        \label{fig:2d}
    \end{subfigure}
    \caption{Distance from the robot to the obstacle boundary over time for (a) MPC-HOCBF, (b) MPC-DHOCBF, (c) MPC-DC, and (d) MPC-TLR under different prediction horizons and safety-constraint parameters. The dashed horizontal line denotes the obstacle boundary, and the insets provide magnified views of the minimum-distance regions.
    } 
    \label{fig:2}
\end{figure*}

\begin{figure*}[!t]
    \vspace*{-0.4cm}
    \centering
    \begin{subfigure}[t]{0.26\linewidth}
        \centering
        \includegraphics[width=1\linewidth]{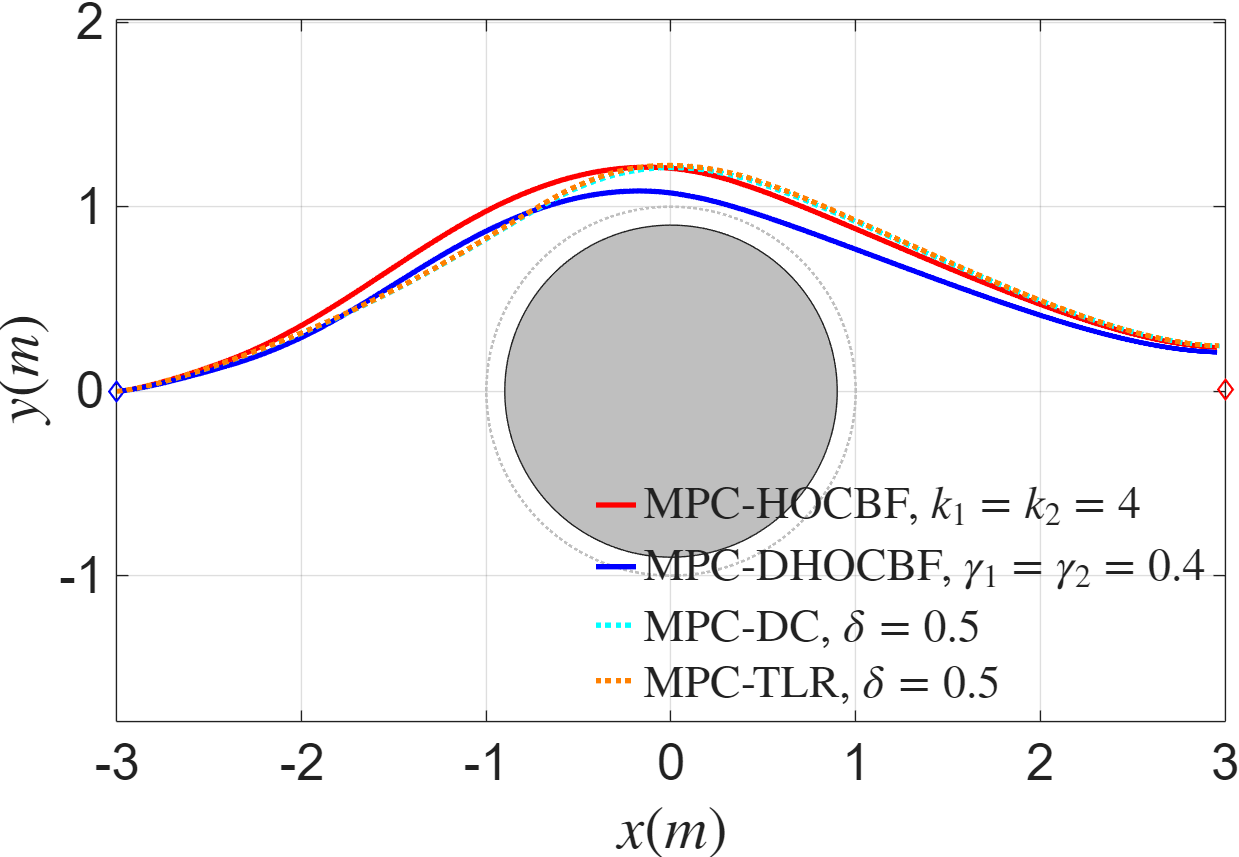}
        \caption{Closed-loop trajectories when $N=24$.}
        \label{fig:3a}
    \end{subfigure}
    \begin{subfigure}[t]{0.26\linewidth}
        \centering
        \includegraphics[width=1\linewidth]{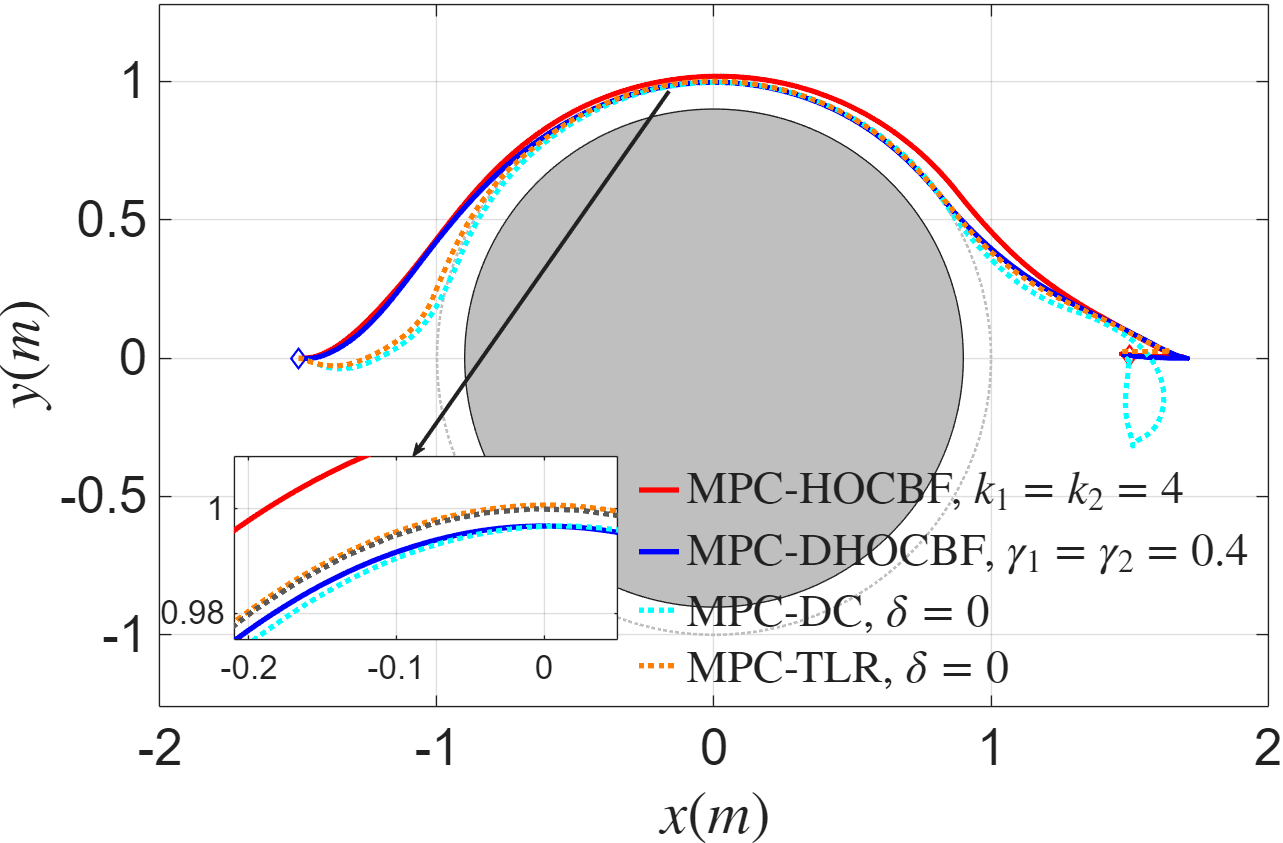}
        \caption{Closed-loop trajectories when $N=4$.}
        \label{fig:3b}
    \end{subfigure}  
    \begin{subfigure}[t]{0.23\linewidth}
        \centering
        \includegraphics[width=1\linewidth]{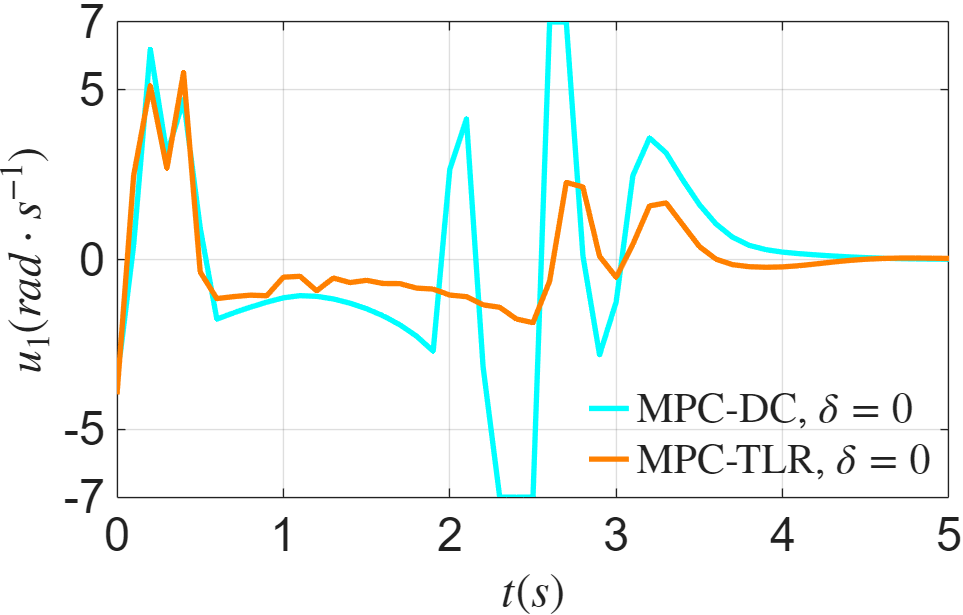}
        \caption{$u_1$ over time when $N=4$.}
        \label{fig:3c}
    \end{subfigure}
    \begin{subfigure}[t]{0.23\linewidth}
        \centering
        \includegraphics[width=1\linewidth]{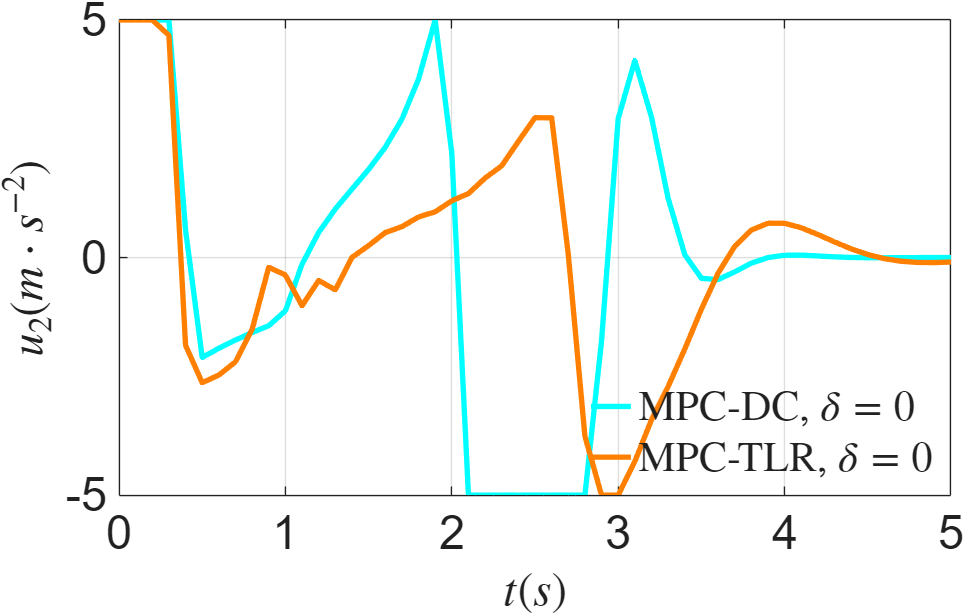}
        \caption{$u_2$ over time when $N=4$.}
        \label{fig:3d}
    \end{subfigure}
    \caption{Comparison of the closed-loop trajectories and control inputs generated by the four safety-critical MPC approaches for prediction horizons $N=24$ and $N=4$. The dashed gray circle indicates the actual obstacle boundary, while the filled gray disk is shown slightly smaller for visual clarity. The inset in (b) provides a magnified view of the inter-sampling trajectory relative to the obstacle boundary.
    } 
    \label{fig:3}
\end{figure*}

Figure~\ref{fig:1} compares the closed-loop trajectories obtained by the four approaches under different prediction horizons and safety-constraint parameters. In general, increasing the prediction horizon allows the controller to anticipate the obstacle earlier, resulting in an earlier initiation of the avoidance maneuver and a smoother closed-loop trajectory. For MPC-HOCBF in Fig.~\ref{fig:1a}, the case with $N=12$ becomes infeasible before the robot reaches the goal, as indicated by the premature termination of the trajectory. In contrast, the longer horizons remain feasible and successfully guide the robot around the obstacle. For MPC-DC in Fig.~\ref{fig:1c}, the case with $N=12$ converges to an undesirable local minimum, preventing the robot from approaching the target state sufficiently closely, whereas increasing $N$ enables the optimizer to identify feasible trajectories around the obstacle. Similar horizon-dependent behavior can also be observed for MPC-DHOCBF and MPC-TLR.

The safety-constraint parameters determine the conservativeness of the resulting trajectories. For MPC-HOCBF and MPC-DHOCBF, smaller values of $k_i$ and $\gamma_i$, respectively, impose more restrictive safety conditions, causing the robot to begin avoiding the obstacle earlier and maintain a larger minimum clearance from its boundary. In contrast, for MPC-DC and MPC-TLR, a larger value of $\delta$ produces a more conservative trajectory and consequently a larger minimum distance from the obstacle. These results illustrate the trade-off between conservativeness, feasibility, and trajectory efficiency introduced by both the prediction horizon and the safety-constraint parameters.

Figure~\ref{fig:2} corresponds to the four trajectory plots in Fig.~\ref{fig:1} and shows the distance from the robot to the obstacle boundary over time. Among the four approaches, MPC-HOCBF maintains the largest clearance and is therefore the safest under the tested settings, but its stronger conservativeness also leads to longer avoidance trajectories. The other three methods exhibit different degrees of inter-sampling safety violation, as indicated by negative minimum distances. As discussed in Sec.~\ref{subsec: Comparative Analysis}, this is caused by the mismatch between the discrete-time prediction model used in the MPC optimization and the continuous-time dynamics used to update the closed-loop trajectory. Among these three methods, MPC-TLR achieves the largest minimum clearance and is therefore relatively safer than MPC-DHOCBF and MPC-DC. One possible explanation, consistent with Remark~\ref{rem: 2}, is that the safety constraints of MPC-TLR and MPC-HOCBF involve more control inputs than those of MPC-DHOCBF and MPC-DC, providing greater flexibility to shape the predicted trajectory for safety. Moreover, while the number of parameters in HOCBF and DHOCBF increases with the relative degree, MPC-TLR adjusts its conservativeness mainly through the single parameter $\delta$, while remaining less conservative than MPC-HOCBF and yielding safer trajectories than MPC-DHOCBF and MPC-DC.

\begin{table*}[]
\centering
\resizebox{0.99\textwidth}{!}{
\begin{tabular}{|cc|cccccc|}
\hline
\multicolumn{2}{|c|}{Approaches} & $N = 4$ & $N = 8$ & $N = 12$ & $N = 16$ & $N = 20$ & $N = 24$ \\ \hline
\multicolumn{1}{|c|}{\multirow{2}{*}{\begin{tabular}[c]{@{}c@{}}MPC-HOCBF\\ ($k_1=k_2=2$)\end{tabular}}} & mean / std (s) & $13.106\pm7.818$ & $4.721\pm6.418$ & $5.683\pm6.848$ & $6.835\pm10.557$ & $5.361\pm6.642$ & $7.930\pm12.183$ \\
\multicolumn{1}{|c|}{} & feas. rate & 93\% & 98\% & 99\%  &  100\% & 99\% & 100\% \\ \hline
\multicolumn{1}{|c|}{\multirow{2}{*}{\begin{tabular}[c]{@{}c@{}}MPC-HOCBF\\ ($k_1=k_2=4)$\end{tabular}}} & mean / std (s) & $15.372\pm7.688$ & $5.275\pm6.703$ & $5.478\pm5.818$ & $5.723\pm6.292$ & $7.082\pm11.742$ & $6.433\pm8.347$ \\
\multicolumn{1}{|c|}{} & feas. rate & 88\% & 100\% & 99\% & 100\% & 99\% & 98\% \\ \hline
\multicolumn{1}{|c|}{\multirow{2}{*}{\begin{tabular}[c]{@{}c@{}}MPC-DHOCBF\\ ($\gamma_1=\gamma_2=0.4$)\end{tabular}}} & mean / std (s) & $3.722\pm2.972$ & $1.090\pm1.255$ & $1.283\pm1.151$ & $1.406\pm1.426$ & $1.596\pm1.333$ & $1.301\pm1.108$ \\
\multicolumn{1}{|c|}{} & feas. rate & 93\% & 99\% & 99\% & 100\% & 100\% & 97\% \\ \hline
\multicolumn{1}{|c|}{\multirow{2}{*}{\begin{tabular}[c]{@{}c@{}}MPC-DHOCBF\\ ($\gamma_1=\gamma_2=0.6$)\end{tabular}}} & mean / std (s) & $4.888\pm3.219$ & $1.171\pm1.762$ & $1.579\pm1.694$ & $1.563\pm1.672$ & $1.433\pm1.654$ & $1.679\pm2.611$ \\
\multicolumn{1}{|c|}{} & feas. rate & 92\% & 100\% & 100\% & 99\% & 99\% & 96\% \\ \hline
\multicolumn{1}{|c|}{\multirow{2}{*}{\begin{tabular}[c]{@{}c@{}}MPC-DC\\ ($\delta=0$)\end{tabular}}} & mean / std (s) & $2.914\pm2.026$ & $0.705\pm0.887$ & $1.079\pm1.067$ & $1.104\pm1.137$ & $1.113\pm0.959$ & $1.303\pm1.107$ \\
\multicolumn{1}{|c|}{} & feas. rate & 92\% & 100\% & 100\%  &  100\% & 100\% & 100\% \\ \hline
\multicolumn{1}{|c|}{\multirow{2}{*}{\begin{tabular}[c]{@{}c@{}}MPC-TLR\\ ($\delta=0$)\end{tabular}}} & mean / std (s) & $12.970\pm8.897$ & $3.576\pm5.047$ & $5.342\pm5.382$ & $8.278\pm11.437$ & $12.704\pm18.325$ & $10.351\pm11.574$ \\
\multicolumn{1}{|c|}{} & feas. rate & 96\% & 99\% & 100\%  &  100\% & 99\% & 100\% \\ \hline
\end{tabular}
}
\caption{Computation time and feasibility rate of the four safety-critical MPC approaches under different prediction horizons and parameter settings. The computation time is reported as the mean $\pm$ standard deviation
over 100 different initial states, where each value corresponds to solving the MPC problem and generating the resulting one-step state transition. The feasibility rate denotes the percentage of these initial states for which the MPC problem was successfully solved.
\label{tab:compuation-time1}}
\end{table*}

Figure~\ref{fig:3a} shows that, with appropriate parameter tuning, all four methods can generate similar closed-loop trajectories. This highlights the practical advantage of MPC-DC and MPC-TLR, which require fewer safety-related parameters to achieve comparable behavior. In Fig.~\ref{fig:3b}, the prediction horizon is reduced to $N=4$, while the position-state weights in $Q$ and $P$ are increased from $10$ to $1000$, i.e., the first two diagonal entries are set to $1000$. Because of the short horizon, all trajectories pass close to the obstacle. However, the inset shows that MPC-HOCBF and MPC-TLR remain outside the dashed obstacle boundary, whereas MPC-DHOCBF and MPC-DC partially enter the unsafe region. This observation is consistent with Remark~\ref{rem: 2}: MPC-HOCBF and MPC-TLR involve more control inputs in their safety constraints, which becomes particularly beneficial when the horizon is comparable to the relative degree. Figures~\ref{fig:3c} and \ref{fig:3d} compare the control inputs of MPC-DC and MPC-TLR for the same case. The MPC-TLR inputs remain smoother and closer to zero, indicating lower control effort and suggesting that involving more control inputs can also improve optimality.

We generated 100 safe initial states outside the obstacle and solved each MPC formulation to obtain a one-step state transition. As shown in Table~\ref{tab:compuation-time1}, for $N\geq 8$, MPC-TLR and MPC-HOCBF generally require longer computation times than MPC-DHOCBF and MPC-DC, which is consistent with their more complex analytical forms discussed in Sec.~\ref{subsec: Comparative Analysis}. Nevertheless, the four methods achieve comparable feasibility rates for these horizons. When $N=4$, MPC-TLR attains the highest observed feasibility rate, which is consistent with Remark~\ref{rem: 2}, since its constraint involves more control inputs than MPC-DHOCBF and MPC-DC and may improve feasibility for short horizons. At the same time, MPC-TLR requires fewer tuning parameters and is less conservative than MPC-HOCBF, leading to a relatively higher feasibility rate.

\section{Conclusion and Future Work}
\label{sec:Conclusion}
This paper proposed an MPC-TLR framework for safety-critical control. The TLR was approximated numerically under zero-order-hold control and incorporated into a receding-horizon MPC formulation with a tunable safety margin. Comparisons with MPC-HOCBF, MPC-DHOCBF, and MPC-DC showed that MPC-TLR achieves safer trajectories than MPC-DHOCBF and MPC-DC under the tested settings, while requiring fewer tuning parameters and being less conservative than MPC-HOCBF.

Future work will focus on guaranteeing inter-sampling safety and investigating computationally efficient implementations of MPC-TLR.
\bibliographystyle{IEEEtran}
\balance
\bibliography{references.bib}

\end{document}